\makeatletter
\@namedef{ver@picins.sty}{9999/99/99}
\makeatother

\documentclass[fleqn]{2023SCGE}
\begin{document}

\ensubject{subject}

\ArticleType{Article}
\Year{2024}
\Month{February}
\Vol{67}
\No{2}
\DOI{https://doi.org/10.1007/s11433-023-2233-2}
\ArtNo{229811}
\ReceiveDate{Aug 22, 2023}
\AcceptDate{Sep 22, 2023}
\OnlineDate{January 8, 2024}

\title{Black holes as the source of dark energy: a stringent test with high-redshift JWST AGNs}

\author[1,2]{Lei Lei}{}%
\author[1,2]{Lei Zu}{}
\author[1,2]{Guan-Wen Yuan}{}
\author[1,2]{Zhao-Qiang Shen}{}%
\author[1,2]{Yi-Ying Wang}{}
\author[1,2]{Yuan-Zhu Wang}{}
\author[3,2]{\\Zhen-Bo Su}{}
\author[3,2]{Wen-Ke Ren}{}
\author[1,2]{Shao-Peng Tang}{}
\author[1,2]{Hao Zhou}{}
\author[1,2]{Chi Zhang}{}
\author[1,2]{Zhi-Ping Jin}{}
\author[1,2]{\\Lei Feng}{fenglei@pmo.ac.cn}
\author[1,2]{Yi-Zhong Fan}{yzfan@pmo.ac.cn}
\author[1,2]{Da-Ming Wei}{}

\AuthorMark{Lei Lei}

\AuthorCitation{Lei Lei, Lei Zu, Guan-Wen Yuan, Zhao-Qiang Shen, Yi-Ying Wang, Yuan-Zhu Wang, Zhen-Bo Su, Wen-Ke Ren, Shao-Peng Tang, Hao Zhou, Chi Zhang, Zhi-Ping Jin, Lei Feng, Yi-Zhong Fan, and Da-Ming Wei}

\address[1]{Key Laboratory of Dark Matter and Space Astronomy, Purple Mountain Observatory, Chinese Academy
of Sciences, Nanjing 210023, China;}
\address[2]{School of Astronomy and Space Science, University of Science and Technology of China, Hefei 230026, China;}
\address[3]{CAS Key Laboratory for Research in Galaxies and Cosmology, Department of Astronomy, University of Science and Technology of China, Hefei, Anhui 230026, China}


\abstract{Studies have proposed that there is evidence for cosmological coupling of black holes (BHs) with an index of $k\approx 3$; hence, BHs serve as the astrophysical source of dark energy. However, the data sample is limited for the redshifts of $\leq 2.5$. In recent years, the James Webb Space Telescope (JWST) has detected many high-redshift active galactic nuclei (AGNs) and quasars.
Among the JWST NIRSpec-/NIRCam-resolved AGNs, three are determined to be in early-type host galaxies with a redshift of $z\sim 4.5--7$. However, their $M_{\star}$ and $M_{\rm BH}$ are in tension with the predicted cosmological coupling of black holes with $k = 3$ at a confidence level of $\sim 2\sigma$, which
challenges the hypothesis that BHs serve as the origin of dark energy.
Future work on high-redshift AGNs using the JWST will further assess such a hypothesis by identifying more early-type host galaxies in the higher mass range.
}



\maketitle


\begin{multicols}{2}
\section{Introduction}\label{section1}

Dark energy is a hypothetical form of energy that acts opposite to gravity to direct the accelerating expansion of the universe  \cite{1998AJ....116.1009R,1999ApJ...517..565P}. Its nature remains a mystery and numerous dark energy models have been proposed in the literature: cosmological constant \cite{1998AJ....116.1009R,1999ApJ...517..565P}, quintessence or k-essence \cite{2021PhRvD.103h1305B,2022JCAP...04..004L}, coupled dark energy \cite{2003PhRvD..68b3514A}, unified dark energy \cite{2004PhRvD..69h3517C} and so on \cite{2012Ap&SS.342..155B}. 
Very recently, there are claims that there is strong evidence for cosmologically coupled mass growth of the black holes (BHs) at redshifts below $2.5$ and stellar remnant BHs are the astrophysical origin of dark energy \cite{2023ApJ...944L..31F,2023ApJ...943..133F}. 
Extensive attention has been paid to such an attractive probability, and dedicated investigations/comments have been published \cite{2023RNAAS...7..101M,2023arXiv230209690M,2023arXiv230401059W,2023arXiv230315793D,Fukuyama:2023lxc,Sadeghi:2023cpd,Cadoni:2023lum,Amendola:2023ays,Adil:2023ara,Gaur:2023hmk}. 
 
BHs are formed in various ways in the universe. Stellar-mass BHs in the range of $\sim 2.2M_{\odot}$ \cite{Han:2022rug,Gao:2023keg} to $\sim10^{2}M_{\odot}$ are the final product of the explosion of massive stars \cite{Remillard2006ARA&A..44...49R}. Intermediate-mass BHs in the range of $\sim10^{2}M_{\odot}$ to $\sim10^{6}M_{\odot}$ are formed through accretion or merger of the BHs \cite{Chilingarian2018ApJ...863....1C, Greene2020ARA&A..58..257G}. Supermassive BHs (SMBHs) with masses exceeding $10^{6}M_{\odot}$ are found as active galactic nuclei (AGNs) or quasars at the center of galaxies in local and high-redshift universes. High-redshift SMBHs in early-type galaxies are classic targets for investigating the theory that cosmologically coupled BHs are sources of dark energy.
If there is indeed a coupling, the mass of the BH is expected to grow with the Robertson–Walker scale factor $a$ as follows \cite{2020ApJ...889..115C}:
\begin{equation} M(a)=M(a_{i})\times\left(\frac{a}{a_{i}}\right)^{k}\qquad  a\geqslant a_{i}, 
\label{eq:1} 
\end{equation} 
where $a$ denotes the scale factor corresponding to lower redshift, $a_{i}$ is the high-redshift BHs, and the index $k$ is the cosmological coupling strength. The cosmological coupling BH dark energy model is a nonstandard model for estimating dark energy. The relation of Eq.~\ref{eq:1} is difficult to justify using standard cosmology.
To serve as dark energy, we would have an index of $k = 3$; therefore, the masses of BHs would have increased by a factor of 343 in early-type galaxies from redshift $z\sim6$ to $z\sim0$. However, the data sample obtained from an earlier study \cite{2023ApJ...944L..31F} is limited to $z\leq 2.5$, and objects at higher redshifts are critical for assessing the hypothesis of the BH growth by cosmological coupling. For such purposes, we need a group of high-redshift AGNs with early-type host galaxies. The excellent performance of the James Webb Space Telescope (JWST) has offered us a valuable opportunity to perform such investigations  \cite{2023MNRAS.518.4755A,2023MNRAS.519.1201A,2023arXiv230712487W}. 
JWST is a powerful infrared telescope that can observe numerous high-redshift galaxies and AGNs. AGNs are powered by their center SMBH accretion. There are luminous broad emission lines in the spectrum of type 1 AGNs with which we can compute the masses of the BHs \cite{2005ApJ...630..122G}.

\section{Sample Selection} \label{sec:sample}
\begin{figure}[H]
\centering
\includegraphics[scale=0.44]{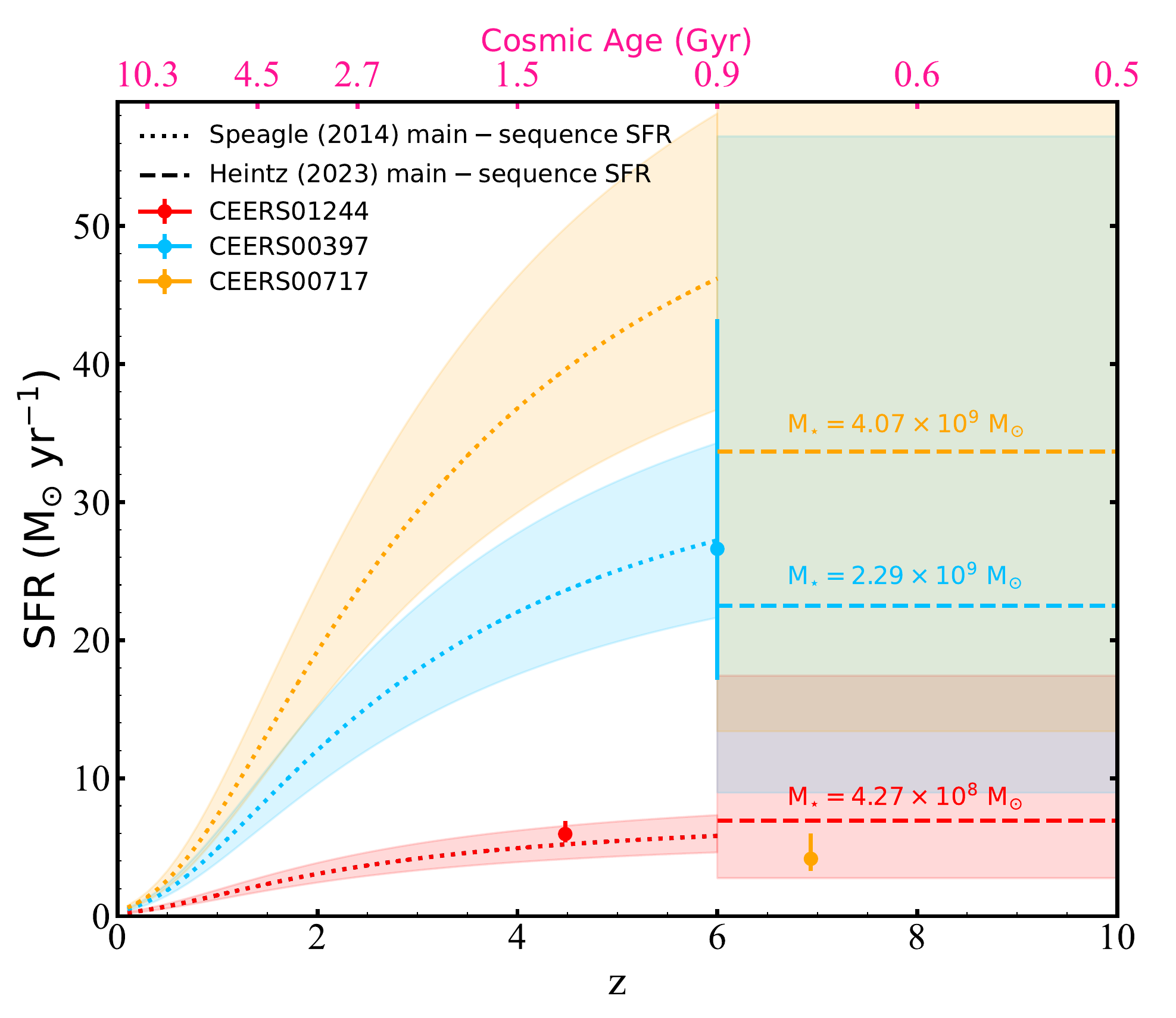}
\caption{The SFRs of selected AGNs. The dotted and dashed lines are "main-sequence" SFR at different redshifts from two references \cite{2014ApJS..214...15S} and \cite{2022arXiv221202890H} respectively. The "main-sequence" SFR curves in red, orange and green correspond to the galaxy stellar mass $M_{\star}$ that is equal to the three selected AGN host galaxies, respectively. 
\label{fig:1}}
\end{figure}

JWST has detected many high-redshift AGNs and quasars. A previous study ~\cite{2023arXiv230311946H} looked for broad-line type-1 AGNs at $z = 4.0--7.0$ among 185 high-redshift galaxies and found a good sample consisting of 10 objects.
To assess the cosmological coupling hypothesis, the host galaxies must be quiescent so that the masses of the BHs would not be improved mainly by accretion or merging.
In this work, we select our high-redshift AGNs with early-type or quiescent host galaxies using the following criteria:
\begin{itemize}
\item Red-shift (z): We select the AGNs at high redshifts of $z>4.0$.

\item H$\alpha$ broad emission line significance (${\rm SNR}_{\rm H \alpha,broad}$) $>5\sigma$:
BH masses were estimated using H$\alpha$ broad emission line of the type-1 AGNs, which is confirmed by Ref.~\cite{2023arXiv230311946H} with JWST/NIRSpec. The AGNs with confident H$\alpha$ broad emission lines are selected to reduce systematic error.

\item Minimally reddened AGNs with ${E(B-V)} < 0.1$: This criterion is the same as that in Ref.~\cite{2023ApJ...944L..31F} for the high redshift samples of HST/COSMOS.

\item Image selection: Following Ref.~\cite{2023ApJ...944L..31F},  we select the AGN host galaxies without unresolved point sources or disks. The images are obtained from Figure 5 of Ref.~\cite{2023arXiv230311946H}.

\item Star formation rate (SFR): we select the AGN host galaxies that are below the $SFR-M_{\star}$ main sequence relationship $\psi_{SFR}$ and the intrinsic scatter $\delta_{\rm SFR}$ at the redshift of the source. The $SFR - M_{\star}$ main sequence relationship at $z\sim 0-6$ is adopted from Ref.~\cite{2014ApJS..214...15S}, while for $z\sim 6-10$ we use the relationship proposed by the JWST  data \cite{2022arXiv221202890H}, i.e., 
\begin{equation}
\log_{10} \psi_{\rm SFR} \left(M_{\odot}/yr \right) =\left\{
\begin{array}{c}
\left(0.96 - 0.045\times t\right)\times \log_{10}M_{\star} \\
- \left(7.41 - 0.27\times t\right),       \\    \delta_{\rm SFR}=0.1 ~{\rm dex},\quad   \\ {0< z \leqslant 6},\\
0.7\times \log_{10}M_{\star}-5.2,      \\    \; \delta_{\rm SFR}=0.4 ~{\rm dex},\quad ~ \\ {6 < z \leqslant 10},
\end{array} \right.
\label{eq:2}
\end{equation}
where $t$ is the cosmic age corresponding to the redshift $z$.

We estimate the SFRs of AGN host galaxies using the H$\alpha$ narrow emission line. Because the broad H$\alpha$ emission line primarily originates from the broad-line region of AGN, which is located near their central SMBHs, while the narrow line emerges from both the narrow line region of AGNs and star formation regions within their host galaxies, the narrow H$\alpha$ emission can be used to trace the star formation activity in AGNs, as it is associated with ionized hydrogen regions where massive stars are formed \cite{Alaghband-Zadeh2016MNRAS.459..999A}. Therefore, we use the luminosity of the H$\alpha$ narrow line to calculate the SFRs \cite{2007arXiv0709.1473P}, i.e., 
\begin{equation}
    {\rm SFR}_{\rm H _{\alpha}} \left(M_{\odot}/yr\right) = 7.9\times10^{-42} \times L_{\rm H \alpha,narrow},
    \label{eq:3}
\end{equation}
where $L_{\rm H \alpha,narrow}$ refers to the luminosity of the narrow emission line of AGN, $\rm erg~s^{-1}$. Figure \ref{fig:1} illustrates the SFR of the selected AGN host galaxies and the SFR criteria calculated using Equation (\ref{eq:2}).
\end{itemize}

The masses of the BHs employed herein were adopted from Ref.~\cite{2023arXiv230311946H}, which were calculated using the $H\alpha$ emission line $M_{BH}=2.0\times10^6M_{\odot}\times(\frac{L_{H\alpha,broad}}{10^{42}\,erg\,s^{-1}})^{0.55}\times(\frac{FWHM_{H\alpha,broad}}{10^3\,km\,s^{-1}})^{2.06}$ (see Equation (16) in H23). The host stellar masses are also obtained from Ref.~\cite{2023arXiv230311946H}, which were estimated by SED fitting after decomposing contributions from the hosts (see Section 4.5 of Ref.~\cite{2023arXiv230311946H}). In addition, a Chabrier initial mass function (IMF) was used in Ref.~\cite{2023arXiv230311946H}. The names of sources, along with their spectroscopic redshift ($z_\mathrm{{spec}}$) and extinction (${E(B-V)}$) values, are obtained from \cite {2023arXiv230311946H} and \cite {2023arXiv230112825N}. Other parameters such as  $L_\mathrm{broad}/L_\mathrm{narrow}$, $L_\mathrm{H\alpha,broad}$, $M_{\rm BH}$, $M_\mathrm{\star}$, and $\lambda_\mathrm{Edd}$ are obtained from Ref.~\cite{2023arXiv230311946H}. The SFR was calculated using Equation (\ref{eq:3}). Table \ref{tab:1} summarizes the physical properties of AGNs and host galaxies for these three sources.

Among the 10 JWST/NIRSpec confirmed type-1 AGNs reported by Ref.~\cite{2023arXiv230311946H}, we select three AGNs with early-type host galaxies: CEERS 01244, CEERS 00397 and CEERS 00717, which are classified as dwarf early-type galaxies (ETGs) due to their lower masses than the samples from Ref.~\cite{2023ApJ...943..133F}, which have masses higher than $4\times 10^{10}\, M_{\odot}$.

Dwarf ETGs are assumed to have been formed in the early universe and have not experienced considerable evolution compared to local low-mass ETGs. Thus, low-mass AGNs hosting ETGs can be observed in high-redshift universes such as those employed here. Evidence for the rapid formation of low-mass ($\sim 10^9 M_{\odot}$) ETGs in the early universe (i.e., at the age of $>10$ Gyr now) was reported by Ref.~\cite{2016ApJ...818..179L}. Among their 11 ETGs within the Virgo Cluster, they identified three old dwarf ETGs (age $>12$ Gyr). Moreover, several studies have identified AGNs residing in low-mass ETGs from local datasets \cite{2015ApJ...813...82R,2017ApJ...840...68G,2018ApJ...868..152B}. More recently, Ref.~\cite{2018ApJ...854....4B} identified some AGNs via long-term optical variability at $z<0.14$. The host galaxy masses for NSA32653 and NSA38720 AGNs are estimated to be $2.94\times 10^9\, M_{\odot}$ and $5.68\times 10^9\, M_{\odot}$, respectively. They also have elliptical host galaxies without any complex structures surrounding their galactic nuclei, as seen in images. Additionally, these two AGNs exhibit good agreement with the local $M_{*}-M_{BH}$ plane, as shown by green stars in Figure \ref{fig:2}.

\begin{table*}
\caption{Physical Properties of Our Selected AGNs and Their Host Galaxies.}
\label{tab:1}
\begin{tabular}{cccccccccccc}

\hline\hline
Name &  $z_{spec}$ &  ${E(B-V)}$ &   $L_\mathrm{H\alpha,broad}$ & $\frac{L_\mathrm{broad}}{L_\mathrm{narrow}}$ &  $M_\mathrm{BH}$ &  $log_{10}(M_\mathrm{\star})$ &  $\lambda_\mathrm{Edd}$ &  $SFR$ \\

& & (mag) &  $erg\, /s$ &  &  $M_{\odot}$  &  $M_{\odot}$  &  &  $M_{\odot}\, /yr$\\
\hline
CEERS01244$^\dagger$ & $4.478$  & $0.00$ & $1.9^{+0.0}_{-0.1}\times10^{42}$ & $2.52^{+0.16}_{-0.23}$ & $1.5^{+0.1}_{-0.1}\times10^{7}$ & $8.63^{+0.63}_{-1.03}$ & $0.30^{+0.56}_{-0.11}$ & $5.96^{+0.95}_{-0.60} $\\
CEERS00397$^\dagger$ & $6.000$  & $0.00$  & $6.4^{+2.1}_{-1.0}\times10^{41}$& $0.19^{+0.07}_{-0.05}$ & $4.9^{+3.6}_{-2.2}\times10^{6}$& $9.36^{+0.36}_{-0.45}$ & $1.31^{+10.89}_{-1.16}$ & $26.61^{+16.67}_{-9.50}$ \\
CEERS00717$^\dagger$ & $6.936$  & $0.00$  & $1.2^{+0.3}_{-0.2}\times10^{41}$& $2.39^{+1.10}_{-0.59}$ & $8.3^{+4.9}_{-2.7}\times10^{6}$& $9.61^{+0.77}_{-1.18}$ & $0.05^{+0.10}_{-0.03}$ & $4.16^{+11.65}_{-5.60}$\\
\hline
\end{tabular}
\\

Note: $^\dagger$:The names and spectroscopic redshifts ($z_{\rm spec}$) are obtained from \cite{2023arXiv230311946H} and \cite{2023arXiv230112825N}. The extinction (${E(B-V)}$), $L_\mathrm{H\alpha,broad}$, $L_\mathrm{broad}/L_\mathrm{narrow}$, $M_{\rm BH}$, $M_\mathrm{\star}$ and $\lambda_\mathrm{Edd}$ are obtained from \cite{2023arXiv230311946H}. The SFR is computed using Equation~(\ref{eq:3}).\\
\end{table*}

\section{Methods} \label{sec:methods}
\subsection{$M_{\star}-M_{\rm BH}$ Fundamental Plane}
SMBHs are believed to lurk at the center of all massive galaxies and play a crucial role in the formation and evolution of the host galaxies \cite{Kormendy2013ARA&A..51..511K,Zhuang+Ho+2023NA}. For instance, the correlations among the mass of the SMBH and properties of its host galaxy (e.g., total stellar mass $M_{*}$, luminosity $L_{\rm{host}}$, and stellar velocity dispersion $\sigma_{*}$.) indicate an evolutionary connection between them, and the tightness correlations between them may manifest as connection between the activity history of central nuclear and galaxy formation \cite{Marconi2003ApJ...589L..21M, Graham2011MNRAS.412.2211G, Beifiori2012MNRAS.419.2497B}. In the channel for the BH mass growth, it can be regarded as a result of not being cosmologically coupled (e.g., via accretion or mergering, see Ref.~\cite{2016PASA...33...32V} and Ref.~\cite{2018ApJ...854....4B}). 
WIthout cosmological coupling, a similar $M_{\star}-M_{\rm BH}$ relation can be observed at different redshifts within the same host galaxy type (e.g., elliptical or ETGs).  Conversely, if the mass of the SMBH is influenced by cosmological coupling, then the $M_{\star}-M_{\rm BH}$ relation of SMBHs varies across different redshifts. Besides, in the case of cosmological coupling, the BH mass will increase even if the galaxies have stopped their growth. The co-evolution or not with the redshift is still a debated \cite{Sun2015ApJ...802...14S, Ding2020ApJ...888...37D}. In this section, we show how JWST AGNs can help to check whether the BH masses are consistent with the cosmological coupling hypothesis or not. 

\begin{figure}[H]
\centering
\includegraphics[scale=0.44]{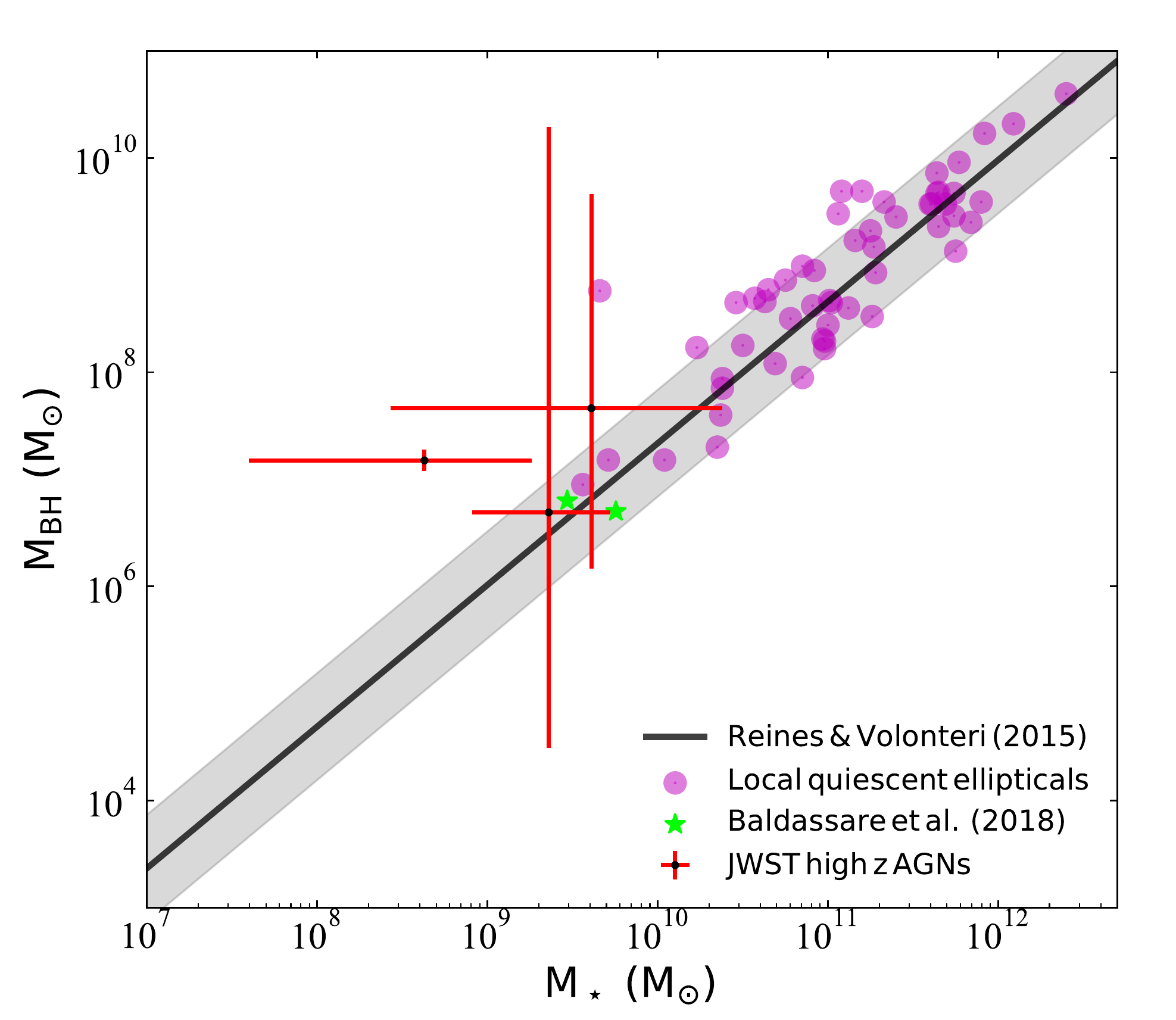}
\caption{Our selected three AGNs and the local sample from Ref.~\cite{2023ApJ...943..133F}. The black line is the $M_{\star}-M_{\rm BH}$ fundamental plane relationship of local elliptical host AGN, the red data points with error bars are the studied AGN here, the purple circles are local quiescent elliptical AGN, and the green stars are two local AGNs in dwarf ETGs obtained from \cite{2018ApJ...854....4B}.
\label{fig:2}}
\end{figure}

Figure \ref{fig:2} presents our selected AGNs in ETGs and the local sample from Ref.~\cite{2023ApJ...943..133F}. The black line refers to the $M_{\star}-M_{\rm BH}$ fundamental plane of local elliptical host AGNs, the red data points with error bars are the three selected AGNs, the purple circles are local quiescent elliptical AGNs, and the green stars are two local AGNs in dwarf ETGs.

We use local quiescent elliptical data from Ref.~\cite{2023ApJ...943..133F} to obtain the black line as follows:
\begin{equation}
    \log_{10} M_{\rm BH} 
    = 8.66_{-0.01-0.5}^{+0.01+0.5}+1.32_{-0.01}^{+0.01}\times \log_{10} \left( \frac{M_{\star}}{10^{11}\ M_{\odot}} \right),
    \label{eq:4}
 \end{equation}
where the errors are the $1\sigma$ standard deviation found in the fitting, and the $0.5\ \rm dex$ refers to the intrinsic scatter of the local sample shown in the gray region in Figure \ref{fig:2}.

One of the sources in our sample has a $M_{\rm BH}$ value higher than that predicted by the cosmological coupling with $k = 3$ by a factor of $\sim10^{3}$, which, hence, sets a very strong constraint on such a possibility. The selected AGNs in this work are ETGs with no more merger or gas-rich images; thus, our sample is typical at that cosmic age.

Note that, for our sample, the $M_*$ is within the range of $\sim 10^8–10^{10}\ M_{\odot}$. In the future, if much more massive high-redshift host galaxies of type-1 AGNs are detected, then the constraint can be examined to the higher stellar mass range.

\subsection{\textbf{Correction of Observational Bias}}
\begin{enumerate}
\setlength{\itemsep}{0pt}
    \item Lauer Selection Bias ($B_{s,vir}$):  \cite{2007ApJ...670..249L} estimated an observational bias for $M_{\rm BH}$, stating that the most massive BHs occur more often as rare outliers in moderate-mass galaxies than in the even rarer high-mass galaxies. The bias should be considered when a flux-limiting survey of quasars observing the luminous or high-mass SMBHs is conducted, especially when mass is higher than $10^{8.5} M_{\odot}$. However, the bias is not important because our data is from one of the deeper fields of the JWST, and our SMBHs have lower mass ($\sim 10^{7} M_{\odot}$) and faint flux (bolometric luminosity $10^{43}\sim 10^{45}L_{\odot}$). Ref.~\cite{2023arXiv230308150Z} computed the $M_{\star}--M_{\rm BH}$ relation at a redshift of $z\sim 6$ (see Fig. 1 in their paper), which indicates that there is only an $\sim-0.1\, \rm dex$ effect for the faint SMBHs ($L_{bol}\leq 10^{45}L_{\odot}$). The range of bias lies between $-0.3\leq B_{s,vir}\leq -0.1\, \rm dex$ in Ref.~\cite{2023ApJ...943..133F}. Hence, we set a selection bias correction limit of $B_{s,vir}\sim -0.3\, \rm dex$ for our faint sample.

    \item Offset of $M_{\rm BH}-L_{\star}$ in high redshift ($B_{L_{\star}}$): The bias is from the cosmological evolution of $M_{\rm BH}/L_{\star}$. In Ref.~\cite{2017MNRAS.472...90D}, the offset was fitted up to $z\sim5$: $\Delta\log M_{\rm BH}=0.5\pm 0.28 \log (1+z)$. The effect of this bias is $\rm \sim -0.42\, dex$ at a redshift of $z\sim6$. Therefore, we set a max selection bias correction $B_{L_{\star}}\sim -0.5\, \rm dex$.
    
    \item Measurement bias ($B_{m,vir}(H\alpha)$): The measurement bias $B_{m,vir}$ is from SMBH mass estimation methods of different emission lines and the selection of the virial factor.
    For the single-epoch $H\beta$ emission line-based SMBH mass estimation, a measurement bias range is $-0.2 \leq B_{m,vir}(H\beta) \leq 0.1\, \rm dex$ by Ref.~\cite{2023ApJ...943..133F}. In this work, we take the BH masses from Ref.~\cite{2023arXiv230311946H} measured by the single-epoch $H\alpha$ line method. We use a correction limit on the BH mass bias of $B_{m,vir}(H\alpha)\sim -0.3 \, \rm dex$. The viral factor adopted for the single-epoch mass measurement is approximately $f_{\rm vir}\approx2.15$, which is lower than the value of $f_{\rm vir}\sim4.5$ adopted by Ref.~\cite{2023ApJ...943..133F}. Thus, there is no bias from the viral factor. Moreover, a bias correction of $B_{m,vir}(H\alpha)\approx -0.3$ is included in this work.

\end{enumerate}

Observational negative biases may lead to overestimation of BH mass.
This would reduce the possibility of finding potential negative evolution in $M_{BH}$ with the redshift, as described by $\frac{M_{BH}}{M_{\star}}=(1+z)^{-k},\, k\approx 3$. Therefore, we have applied negative bias corrections to consider possible impacts from observations and selections. The correction values assumed in this work are summarized in Table \ref{tab:2}. The total bias correction is
\begin{equation}
    B_{BH} 
    = B_{s,vir}+B_{L_{\star}}+B_{m,vir}(H\alpha).
    \label{eq:5}
 \end{equation}

\begin{table*}[htbp]
\centering
\setlength{\tabcolsep}{.5em}
\caption{Bias Corrections.}
\begin{tabular}{c c c c}
\hline
\hline
  & $B_{s,vir}$ &  $B_{L_{\star}}$ & $B_{m,vir}(H\alpha)$\\
\hline
Parameter (dex) &  -0.3  &  -0.5 & -0.3\\
\hline
\end{tabular}
\label{tab:2}
\end{table*}

Here, we only include negative biases because the existence of negative biases results in a lower cosmological coupling index $k$. We correct the negative biases to provide a conservative test in the case of BH as a dark energy source while coupling with the scale expansion of cosmology. The positive biases are converse to the cosmological coupling evolution phenomenon. If positive biases such as “dynamical SMBH bias” and “BH early accretion growth” are considered, $k$ will be lower. Thus, positive biases are not included in our analysis, and discussion can be found in Section \ref{sect:5.1}.

\subsection{Estimation of Parameter $k$}
To estimate the parameter $k$, we combine Eq.~(\ref{eq:1}) and Eq.~(\ref{eq:4}) and have 
\begin{equation}
\begin{array}{c}
\mu_{i}(k) = \log_{10} \left(M_{BH,i}\right) 
=1.32\times\log_{10}M_{\star,i} \\ -k\times \log_{10}(1+z_{i})-5.86.
    \label{eq:6}
\end{array}
\end{equation}
Its likelihood reads
\begin{equation}
\mathcal{L}\left(d \vert k\right) = \prod_i \frac{1}{\sigma_{i}\,\sqrt{2\pi}} \exp \left[ -\frac{\left(\mu_{i}\left(k\right)- \left(\lambda_{i} + B_{BH} \right)\right)^2}{2(\sigma_{i})^2} \right],
\label{eq:7}
\end{equation}
where $\lambda_{i}$ and $\sigma_{i}$ are the observed BH mass and the error including the uncertainties of $M_{{\rm BH,}i}$ and $M_{\star,i}$, respectively. 
$B_{BH}$ is the total bias correction of the $M_{BH}$ from Eq.~(\ref{eq:5}).
Also, the total error is the sum of measured BH mass error ($\sigma_{BH,i}$), conversion error ($\sigma_{{\rm conv},i}$) from stellar mass and fundamental plane intrinsic scatter $\sigma_{\rm intrinsic}=0.51\;\rm dex$:
\begin{equation}
\sigma^2_{{\rm tot},i}=\sigma^2_{BH,i}+\sigma^2_{{\rm conv},i}+\sigma^2_{\rm intrinsic}.
\label{eq:8}
\end{equation}
The errors in our data are in the logarithmic space. 
Afterward, we can fit the posterior distribution of $k$ by the likelihood.

\section{Result} \label{sec:result}
\begin{figure}[H]
\includegraphics[scale=0.37]{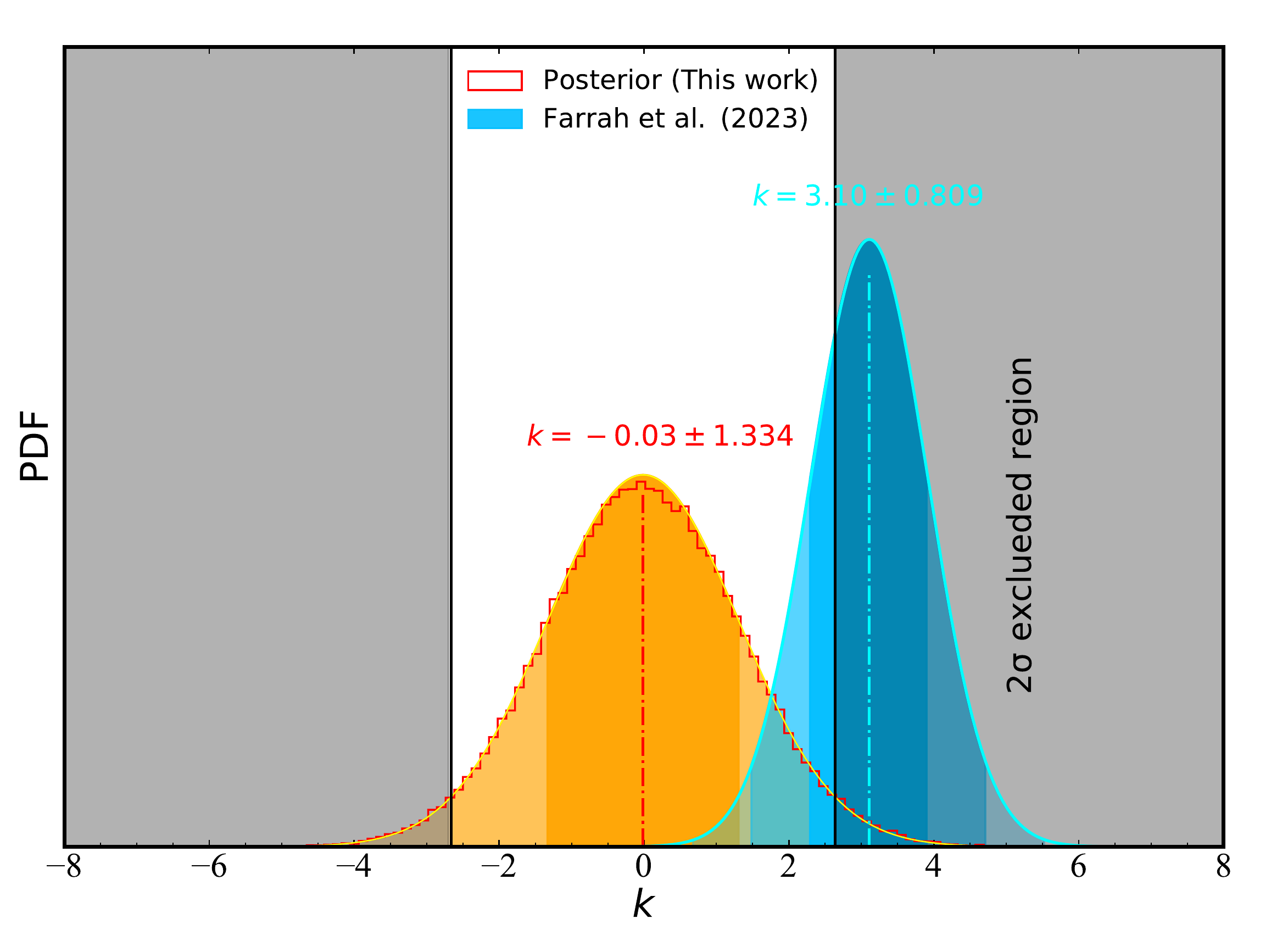}
\caption{Probability distribution function of the posterior of parameter $k$ and our constraint on the hypothesis that the SMBHs are the astrophysical origin of dark energy. The black line represents our $2\sigma$ upper bound on the cosmological coupling index $k$, which is strongly in tension with $k=3$ that is needed to serve as the dark energy. The orange dashed and cyan lines denote the most probable values of $k$ fitted with our sample and that of Ref.~\cite{2023ApJ...944L..31F}, respectively.  
\label{fig:3}}
\end{figure}

As described in Section \ref{sec:methods} and shown in Figure \ref{fig:2}, JWST AGNs with early-type host galaxies have much higher BH mass than that predicted using cosmological coupling growth theory.
Figure \ref{fig:3} illustrates the constraint on the coupling index $k$ (the yellow region).
The best-fit value is $k=-0.03\pm1.334$ (68\% confidence interval), which deviates from the "dark energy hypothesis" (i.e., $k = 3$) at a confidence level of $>2\sigma$; thus, it challenges the claim that the stellar-remnant BHs are the astrophysical sources of dark energy.

Ref.~\cite{2023ApJ...944L..31F} employed $\sim 440$ SMBHs in ETGs at $z\sim0.8$ and $\sim20$ SMBHs in ETGs up to $z\sim 2.5$. In this work, we use three type-1 AGNs at $z\sim6$ in ETGs, which are selected from 10 type-1 AGNs of $\sim 180$ JWST/NIRSpec confirmed galaxies. Although the stellar masses of our sample are lower than most of the samples from another study \cite{2023ApJ...944L..31F}, overlap with the local quiescent ellipticals can be observed.

\section{Discussion}
Recently, the evidence for cosmological coupling of BHs with an index of $k\approx 3$ has been proposed. If confirmed, this likely indicates the astrophysical source of dark energy. The data sample presented in Ref.~\cite{2023ApJ...944L..31F}, however, is limited for the redshifts of $\leq 2.5$. Due to the excellent performance of the JWST, a few hundred high-redshift AGNs and quasars have been detected.
Three JWST NIRSpec-/NIRCam-resolved AGNs have been found within early-type host galaxies with redshifts of $z\sim 4.5--7$. Their $M_{\star}$ and $M_{\rm BH}$, however, are ($\geq 2\sigma$) in tension with the predicted cosmological coupling of BHs with $k = 3$.
Therefore, this challenges the claim that BHs serve as the origin of dark energy (see also Ref.~\cite{2023ApJ...947L..12R} and Ref.~\cite{2023arXiv230501307A} for independent observational tests for the hypothesis of cosmological coupling of BHs).
The $k=0$ case is still allowed at the $\sim 1\sigma$ confidence level, implying that the BHs may be just the "normal" matters in the universe. 

The growth history of BHs is an important but complicated challenge in astronomy and cosmology. The SMBHs in the high-redshift universe can grow with accretion or merger processes \cite{2016PASA...33...32V,2018ApJ...854....4B}. 
There are even arguments that the primordial BHs with an initial mass of $\sim 100M_\odot$ can grow up to SMBHs by very efficient accretion in the early universe \cite{2023arXiv230309391Y,2023arXiv230408153S}.
Before the JWST, there were some analyses on the AGN co-evolution and SMBH mass at high redshifts, but the details are still unclear \cite{2016PASA...33...32V,2021IAUS..356..261T}. The JWST is a powerful tool to observe the high-redshift universe, as demonstrated in  Ref.~\cite{2023arXiv230311946H} in identifying faint type-1 AGNs and their host galaxies.
Future observations of the JWST will be detecting a large number of high-redshift faint AGNs, allowing investigation of the galaxy and SMBH co-evolution or $M_{\star}-M_{\rm BH}$ fundamental plane in the early universe. Also, the JWST may find some AGNs with very massive host galaxies in the early universe, which will further test the hypothesis that the SMBHs serve as the astrophysical origin of dark energy.

\subsection{Positive Bias}
\label{sect:5.1}
Positive bias correction can reduce the possible cosmological coupling strength $k$, but it is not useful in examining the potential coupling growth of BHs. Hence, we list these biases below but do not use them in our analysis.
\begin{enumerate}
\setlength{\itemsep}{0pt}
    
    \item Stellar-Mass Bias ($B_{m,M_{\star}}$): The $B_{m,M_{\star}}$ bias is based on the differences exhibited by methods for estimating the stellar mass of hosts and passive stellar evolution between the two epochs. Ref.~\cite{2015ApJ...813...82R} reported stellar masses that are $0.33\rm\, dex$ lower than those reported in another work\cite{2021ApJ...907....6Z}. Thus, Ref.~\cite{2023ApJ...943..133F} suggested the bias limit to be $\sim0.33\rm\, dex$.
    However, in this work, the stellar masses adopted are from Ref.~\cite{2023arXiv230311946H}, which estimated the host stellar contribution of the image by decomposing the host and AGN using the point spread function (PSF)+Sersic profile fitting method. The stellar masses are computed using the SED fitting with the assumed Chabrier IMF. Ref.~\cite{2023ApJ...943..133F} suggested a lower limit $\sim0.08 \rm dex$ for this bias because they used the Kroupa IMF and scaled it to the Chabrier IMF. Therefore, the bias range of our selection is $0.0\leq B_{m,M_{\star}} \leq 0.33\, \rm dex$. 

    \item Early Accretion Growth Bias ($B_{g}$): The $B_{g}$ bias emerges because the observed SMBHs have not completed their accretion phase, and their masses will grow after the observed states. The bias range of our AGNs is $0.0\leq B_{g} \leq 1.0\, \rm dex$. 

    \item Dynamical SMBH bias ($B_{s,dyn}$): The $B_{s,dyn}$ bias arises because observed dynamical velocity dispersions may be higher than those observed in the near fields of SMBHs, resulting in the estimation of larger SMBH mass. Based on Ref.~\cite{2020NatAs...4..282S} and Ref.~\cite{2023ApJ...943..133F}, our AGN sample has a bias range of $0.0\leq B_{g} \leq 1.3\, \rm dex$ in terms of this effect.
        
\end{enumerate}

\subsection{Possible Explanations for the $M_{\rm BH}-M_{\star}$ Redshift Evolution Phenomenon at $k\approx 3$}
\begin{enumerate}
\setlength{\itemsep}{0pt}
    
    \item S/S0 host galaxies: Because of the PSF, even space observations are not clear enough to distinguish elliptical (E) or spiral (S/S0) galaxies at high redshifts; this may result in the pollution of the spiral host galaxies. Ref.~\cite{2016ApJ...831..134V} and \cite{2023MNRAS.521.1023G} explored the $M_{\rm BH}--M_{\star}$ relation in different types of host galaxies and revealed that AGNs in S/S0 host galaxies have lower BH masses (see Fig. 2 in Ref.~\cite{2016ApJ...831..134V} and Fig. 2 in Ref.~\cite{2023MNRAS.521.1023G}).

    \item Obscured AGNs: Ref.~\cite{2010A&A...522L...3S} studied the $M_{\rm BH}--M_{\star}$ relationship with selected obscured AGNs in high redshifts. The obscured AGNs exhibit obvious lower BH masses in this relation (see Fig. 2 in their paper). The obscured AGNs are difficult to exclude while investigating morphology; hence, the large AGN in the elliptical host galaxy sample may have mixed any obscured AGNs at high redshifts. 

    \item Star-Forming (SF) host galaxies: Ref.~\cite{2023A&A...672A..98M} differentiated the SF-dominant and AGN-dominant host galaxies and fitted the $M_{\rm BH}--M_{\star}$ relation. The SMBHs in SF host galaxies have lower masses than those in AGN host galaxies (see Fig. 6 in their paper). It is difficult to distinguish between the SF or AGN host galaxies in high redshifts; therefore, the large sample may be polluted with SMBHs in the SF host.  
        
\end{enumerate}

\subsection{Future Observations of High Redshift AGNs $M_{\rm BH}-M_{\star}$ Relation}
The spatial resolution and sensitivity of a telescope are crucial for investigating the morphologies of AGN host galaxies. SDSS (\cite{Kent+etal+1994}; \cite{Fukugita+etal+1996}), Pan-STARRS (\cite{Jedicke+etal+2007}; \cite{Chambers+etal+2016}), SkyMapper (\cite{Schmidt+etal+2005}; \cite{Rakich+etal+2006}), DES(\cite{2005astro.ph.10346T}; \cite{2016MNRAS.460.1270D}) and other built telescopes have obtained numerous observation data but are not powerful enough to explore the morphologies of AGN host galaxies. 
Soon, survey telescopes such as LSST (\cite{Hlozek+etal+2019}), WFST (\cite{Lou+etal+2016}; \cite{Lei+etal+2023}; \cite{WFSTscience}) and Mephisto (\cite{Liu+etal+2019}; \cite{Lei+etal+2020}; \cite{Yuan+etal+2020}; \cite{Lei+etal+2022}) will join their peers and perform deeper multiband surveys to provide photometric data with high spatial resolution. The China Space Station Telescope (CSST; \cite{Zhao+etal+2016}; \cite{Yuan+etal+2021}), ROMAN space telescope (\cite{McEnery+etal+2021}) and other space telescopes will substantially uncover the morphologies of AGN host galaxies. Selection of obscured or SF host AGNs is also challenging, except when multiwavelength data are used, including X-ray, sub-millimeter, infrared, and optical bands. The launched eROSITA X-ray survey telescope \cite{2006SPIE.6266E..0PP,2006SPIE.6266E..0OP} and upcoming Einstein Probe X-ray survey telescope \cite{2022hxga.book...86Y,2015arXiv150607735Y} will provide more scientific production to study the obscured AGNs in combination with the JWST and other telescopes. 

\section{Conclusion and Outlook}

We have studied the growth of SMBH masses in dwarf ETGs and the potential cosmological coupling effect. From high redshift AGNs identified by JWST/NIRSpec
\cite{2023arXiv230311946H}, we found three AGNs in dwarf ETGs. The potential BH cosmological coupling effect was tested through the $M_{\star}-M_{BH}$ relation, where $M_{\star}$ represents host galaxy stellar mass and $M_{BH}$ represents BH mass measured with broad $\rm H\alpha$ emission line. Unfortunately, even after correcting for potential biases, we discovered that the current JWST AGNs are in tension with the BH cosmological coupling effect at a(or 95\%) confidence level. This challenges the theory of cosmological coupled BHs as the source of dark energy.
To address this, we corrected potential negative observational biases that could weaken the cosmological coupling effect. However, there may still be some positive biases, such as the early accretion growth bias, which is not a cosmological coupling effect but could lead to results that resemble a cosmological coupling phenomenon. Furthermore, the type of SMBH host is an important factor in testing the cosmological coupling hypothesis.
The $M_{\star}-M_{BH}$ relations differ between obscured AGNs and unobscured AGNs. However, it is challenging to differentiate between the types of high redshift AGN obscuring materials in a survey limited by flux and spatial resolution. The complex physical processes need to be solved in the future. Additionally, the theory of BHs as the source of dark energy still requires further testing in the future.

Moreover, there are newly built or proposed telescopes that offer opportunities to uncover the evolutionary history of AGNs and their hosts.  Future observations of high redshift AGNs will cover a wider range of host stellar masses, allowing us to test the cosmological coupling effect with SMBHs in massive ETGs.

\Acknowledgements{We thank anonymous referees for their valuable suggestions and comments. We thank Xiang Li and Xian-Zhong Zheng for the helpful discussions. Lei Lei gratefully acknowledges the support of Cyrus Chun Ying Tang Foundations. This work is based on the NASA/ESA/CSA James Webb Space Telescope observation results from Yuichi Harikane's research. This work was supported by NSFC (11921003 and 12233011) and by the Chinese Academy of Sciences via the Key Research Program of Frontier Sciences (No. QYZDJ-SSW-SYS024). Software: astropy \cite{2013A&A...558A..33A,2018AJ....156..123A}, emcee \cite{2013PASP..125..306F}, matplotlib \cite{Hunter:2007}, numpy \cite{harris2020array} }

\InterestConflict{The authors declare that they have no conflict of interest.}


\section{References}\label{sec:8}

\bibliographystyle{scpma}
\bibliography{biblio.bib}

\begin{thebibliography}{10}
\providecommand{\url}[1]{\texttt{#1}}
\providecommand{\urlprefix}{URL }
\providecommand{\doi}[1]{doi:~\href{http://doi.org/#1}{\nolinkurl{#1}}}
\providecommand{\arXiv}[1]{\href{https://arxiv.org/abs/#1}{\nolinkurl{https://arxiv.org/abs/#1}}}
\providecommand{\eprint}[1]{\href{http://arxiv.org/abs/#1}{\nolinkurl{#1}}}

\bibitem{1998AJ....116.1009R}
A.~G. {Riess}, A.~V. {Filippenko}, P.~{Challis}, A.~{Clocchiatti},
  A.~{Diercks}, P.~M. {Garnavich}, R.~L. {Gilliland}, C.~J. {Hogan}, S.~{Jha},
  R.~P. {Kirshner}, B.~{Leibundgut}, M.~M. {Phillips}, D.~{Reiss}, B.~P.
  {Schmidt}, R.~A. {Schommer}, R.~C. {Smith}, J.~{Spyromilio}, C.~{Stubbs},
  N.~B. {Suntzeff}, and J.~{Tonry}, Astron. J. \textbf{116}, 1009 (1998),
  arXiv: \eprint{astro-ph/9805201}.

\bibitem{1999ApJ...517..565P}
S.~{Perlmutter}, G.~{Aldering}, G.~{Goldhaber}, R.~A. {Knop}, P.~{Nugent},
  P.~G. {Castro}, S.~{Deustua}, S.~{Fabbro}, A.~{Goobar}, D.~E. {Groom}, I.~M.
  {Hook}, A.~G. {Kim}, M.~Y. {Kim}, J.~C. {Lee}, N.~J. {Nunes}, R.~{Pain},
  C.~R. {Pennypacker}, R.~{Quimby}, C.~{Lidman}, R.~S. {Ellis}, M.~{Irwin},
  R.~G. {McMahon}, P.~{Ruiz-Lapuente}, N.~{Walton}, B.~{Schaefer}, B.~J.
  {Boyle}, A.~V. {Filippenko}, T.~{Matheson}, A.~S. {Fruchter}, N.~{Panagia},
  H.~J.~M. {Newberg}, W.~J. {Couch}, and T.~S.~C. {Project}, ApJ \textbf{517},
  565 (1999), arXiv: \eprint{astro-ph/9812133}.

\bibitem{2021PhRvD.103h1305B}
A.~{Banerjee}, H.~{Cai}, L.~{Heisenberg}, E.~{\'O}. {Colg{\'a}in}, M.~M.
  {Sheikh-Jabbari}, and T.~{Yang}, PRD \textbf{103}, L081305 (2021), arXiv:
  \eprint{2006.00244}.

\bibitem{2022JCAP...04..004L}
B.-H. {Lee}, W.~{Lee}, E.~{{\'O} Colg{\'a}in}, M.~M. {Sheikh-Jabbari}, and
  S.~{Thakur}, JCAP \textbf{2022}, 004 (2022), arXiv: \eprint{2202.03906}.

\bibitem{2003PhRvD..68b3514A}
L.~{Amendola} and C.~{Quercellini}, PRD \textbf{68}, 023514 (2003), arXiv:
  \eprint{astro-ph/0303228}.

\bibitem{2004PhRvD..69h3517C}
V.~F. {Cardone}, A.~{Troisi}, and S.~{Capozziello}, PRD \textbf{69}, 083517
  (2004), arXiv: \eprint{astro-ph/0402228}.

\bibitem{2012Ap&SS.342..155B}
K.~{Bamba}, S.~{Capozziello}, S.~{Nojiri}, and S.~D. {Odintsov}, Ap\&SS
  \textbf{342}, 155 (2012), arXiv: \eprint{1205.3421}.

\bibitem{2023ApJ...944L..31F}
D.~{Farrah}, K.~S. {Croker}, M.~{Zevin}, G.~{Tarl{\'e}}, V.~{Faraoni},
  S.~{Petty}, J.~{Afonso}, N.~{Fernandez}, K.~A. {Nishimura}, C.~{Pearson},
  L.~{Wang}, D.~L. {Clements}, A.~{Efstathiou}, E.~{Hatziminaoglou}, M.~{Lacy},
  C.~{McPartland}, L.~K. {Pitchford}, N.~{Sakai}, and J.~{Weiner}, ApJl
  \textbf{944}, L31 (2023), arXiv: \eprint{2302.07878}.

\bibitem{2023ApJ...943..133F}
D.~{Farrah}, S.~{Petty}, K.~S. {Croker}, G.~{Tarl{\'e}}, M.~{Zevin},
  E.~{Hatziminaoglou}, F.~{Shankar}, L.~{Wang}, D.~L. {Clements},
  A.~{Efstathiou}, M.~{Lacy}, K.~A. {Nishimura}, J.~{Afonso}, C.~{Pearson}, and
  L.~K. {Pitchford}, ApJ \textbf{943}, 133 (2023), arXiv: \eprint{2212.06854}.

\bibitem{2023RNAAS...7..101M}
T.~{Mistele}, Research Notes of the American Astronomical Society \textbf{7},
  101 (2023), arXiv: \eprint{2304.09817}.

\bibitem{2023arXiv230209690M}
E.~{Mottola}, arXiv e-prints arXiv:2302.09690 (2023), arXiv:
  \eprint{2302.09690}.

\bibitem{2023arXiv230401059W}
Y.~{Wang} and Z.~{Wang}, arXiv e-prints arXiv:2304.01059 (2023), arXiv:
  \eprint{2304.01059}.

\bibitem{2023arXiv230315793D}
P.~K. {Dahal}, F.~{Simovic}, I.~{Soranidis}, and D.~R. {Terno}, arXiv e-prints
  arXiv:2303.15793 (2023), arXiv: \eprint{2303.15793}.

\bibitem{Fukuyama:2023lxc}
T.~Fukuyama  (2023), arXiv: \eprint{2301.09220}.

\bibitem{Sadeghi:2023cpd}
J.~Sadeghi, S.~Noori~Gashti, M.~R. Alipour, and M.~A.~S. Afshar  (2023), arXiv:
  \eprint{2305.12545}.

\bibitem{Cadoni:2023lum}
M.~Cadoni, A.~P. Sanna, M.~Pitzalis, B.~Banerjee, R.~Murgia, N.~Hazra, and
  M.~Branchesi  (2023), arXiv: \eprint{2306.11588}.

\bibitem{Amendola:2023ays}
L.~Amendola, D.~C. Rodrigues, S.~Kumar, and M.~Quartin  (2023), arXiv:
  \eprint{2307.02474}.

\bibitem{Adil:2023ara}
S.~A. Adil, U.~Mukhopadhyay, A.~A. Sen, and S.~Vagnozzi  (2023), arXiv:
  \eprint{2307.12763}.

\bibitem{Gaur:2023hmk}
R.~Gaur and M.~Visser  (2023), arXiv: \eprint{2308.07374}.

\bibitem{Han:2022rug}
M.-Z. Han, Y.-J. Huang, S.-P. Tang, and Y.-Z. Fan, Sci. Bull. \textbf{68}, 913
  (2023), arXiv: \eprint{2207.13613}.

\bibitem{Gao:2023keg}
S.-J. Gao and X.-D. Li  (2023), arXiv: \eprint{2307.10708}.

\bibitem{Remillard2006ARA&A..44...49R}
R.~A. {Remillard} and J.~E. {McClintock}, ARA\&A \textbf{44}, 49 (2006), arXiv:
  \eprint{astro-ph/0606352}.

\bibitem{Chilingarian2018ApJ...863....1C}
I.~V. {Chilingarian}, I.~Y. {Katkov}, I.~Y. {Zolotukhin}, K.~A. {Grishin},
  Y.~{Beletsky}, K.~{Boutsia}, and D.~J. {Osip}, ApJ \textbf{863}, 1 (2018),
  arXiv: \eprint{1805.01467}.

\bibitem{Greene2020ARA&A..58..257G}
J.~E. {Greene}, J.~{Strader}, and L.~C. {Ho}, ARA\&A \textbf{58}, 257 (2020),
  arXiv: \eprint{1911.09678}.

\bibitem{2020ApJ...889..115C}
K.~S. {Croker}, K.~A. {Nishimura}, and D.~{Farrah}, ApJ \textbf{889}, 115
  (2020), arXiv: \eprint{1904.03781}.

\bibitem{2023MNRAS.518.4755A}
N.~J. {Adams}, C.~J. {Conselice}, L.~{Ferreira}, D.~{Austin}, J.~A.~A.
  {Trussler}, I.~{Juod{\v{z}}balis}, S.~M. {Wilkins}, J.~{Caruana}, P.~{Dayal},
  A.~{Verma}, and A.~P. {Vijayan}, MNRAS \textbf{518}, 4755 (2023), arXiv:
  \eprint{2207.11217}.

\bibitem{2023MNRAS.519.1201A}
H.~{Atek}, M.~{Shuntov}, L.~J. {Furtak}, J.~{Richard}, J.-P. {Kneib},
  G.~{Mahler}, A.~{Zitrin}, H.~J. {McCracken}, S.~{Charlot}, J.~{Chevallard},
  and I.~{Chemerynska}, MNRAS \textbf{519}, 1201 (2023), arXiv:
  \eprint{2207.12338}.

\bibitem{2023arXiv230712487W}
Y.-Y. {Wang}, L.~{Lei}, G.-W. {Yuan}, and Y.-Z. {Fan}, arXiv e-prints
  arXiv:2307.12487 (2023), arXiv: \eprint{2307.12487}.

\bibitem{2005ApJ...630..122G}
J.~E. {Greene} and L.~C. {Ho}, ApJ \textbf{630}, 122 (2005), arXiv:
  \eprint{astro-ph/0508335}.

\bibitem{2023arXiv230311946H}
Y.~{Harikane}, Y.~{Zhang}, K.~{Nakajima}, M.~{Ouchi}, Y.~{Isobe}, Y.~{Ono},
  S.~{Hatano}, Y.~{Xu}, and H.~{Umeda}, arXiv e-prints arXiv:2303.11946 (2023),
  arXiv: \eprint{2303.11946}.

\bibitem{2014ApJS..214...15S}
J.~S. {Speagle}, C.~L. {Steinhardt}, P.~L. {Capak}, and J.~D. {Silverman}, ApJs
  \textbf{214}, 15 (2014), arXiv: \eprint{1405.2041}.

\bibitem{2022arXiv221202890H}
K.~E. {Heintz}, G.~B. {Brammer}, C.~{Gim{\'e}nez-Arteaga}, V.~B. {Strait},
  C.~d.~P. {Lagos}, A.~P. {Vijayan}, J.~{Matthee}, D.~{Watson}, C.~A. {Mason},
  A.~{Hutter}, S.~{Toft}, J.~P.~U. {Fynbo}, and P.~A. {Oesch}, arXiv e-prints
  arXiv:2212.02890 (2022), arXiv: \eprint{2212.02890}.

\bibitem{Alaghband-Zadeh2016MNRAS.459..999A}
S.~{Alaghband-Zadeh}, M.~{Banerji}, P.~C. {Hewett}, and R.~G. {McMahon}, MNRAS
  \textbf{459}, 999 (2016), arXiv: \eprint{1603.06419}.

\bibitem{2007arXiv0709.1473P}
J.~{Pflamm-Altenburg}, C.~{Weidner}, and P.~{Kroupa}, arXiv e-prints
  arXiv:0709.1473 (2007), arXiv: \eprint{0709.1473}.

\bibitem{2023arXiv230112825N}
K.~{Nakajima}, M.~{Ouchi}, Y.~{Isobe}, Y.~{Harikane}, Y.~{Zhang}, Y.~{Ono},
  H.~{Umeda}, and M.~{Oguri}, arXiv e-prints arXiv:2301.12825 (2023), arXiv:
  \eprint{2301.12825}.

\bibitem{2016ApJ...818..179L}
Y.~{Liu}, E.~W. {Peng}, J.~{Blakeslee}, P.~{C{\^o}t{\'e}}, L.~{Ferrarese},
  A.~{Jord{\'a}n}, T.~H. {Puzia}, E.~{Toloba}, and H.-X. {Zhang}, ApJ
  \textbf{818}, 179 (2016), arXiv: \eprint{1512.00253}.

\bibitem{2015ApJ...813...82R}
A.~E. {Reines} and M.~{Volonteri}, ApJ \textbf{813}, 82 (2015), arXiv:
  \eprint{1508.06274}.

\bibitem{2017ApJ...840...68G}
A.~W. {Graham}, J.~{Janz}, S.~J. {Penny}, I.~V. {Chilingarian}, B.~C.
  {Ciambur}, D.~A. {Forbes}, and R.~L. {Davies}, ApJ \textbf{840}, 68 (2017),
  arXiv: \eprint{1705.03587}.

\bibitem{2018ApJ...868..152B}
V.~F. {Baldassare}, M.~{Geha}, and J.~{Greene}, ApJ \textbf{868}, 152 (2018),
  arXiv: \eprint{1808.09578}.

\bibitem{2018ApJ...854....4B}
S.~{Ben-Ami}, A.~{Vikhlinin}, and A.~{Loeb}, ApJ \textbf{854}, 4 (2018), arXiv:
  \eprint{1712.03207}.

\bibitem{Kormendy2013ARA&A..51..511K}
J.~{Kormendy} and L.~C. {Ho}, ARA\&A \textbf{51}, 511 (2013), arXiv:
  \eprint{1304.7762}.

\bibitem{Zhuang+Ho+2023NA}
M.-Y. Zhuang and L.~C. Ho, Nature Astronomy  (2023),
  \urlprefix\url{http://dx.doi.org/10.1038/s41550-023-02051-4}.

\bibitem{Marconi2003ApJ...589L..21M}
A.~{Marconi} and L.~K. {Hunt}, ApJl \textbf{589}, L21 (2003), arXiv:
  \eprint{astro-ph/0304274}.

\bibitem{Graham2011MNRAS.412.2211G}
A.~W. {Graham}, C.~A. {Onken}, E.~{Athanassoula}, and F.~{Combes}, MNRAS
  \textbf{412}, 2211 (2011), arXiv: \eprint{1007.3834}.

\bibitem{Beifiori2012MNRAS.419.2497B}
A.~{Beifiori}, S.~{Courteau}, E.~M. {Corsini}, and Y.~{Zhu}, MNRAS
  \textbf{419}, 2497 (2012), arXiv: \eprint{1109.6265}.

\bibitem{2016PASA...33...32V}
R.~{Valiante}, R.~{Schneider}, and M.~{Volonteri}, PASA \textbf{33}, e032
  (2016).

\bibitem{Sun2015ApJ...802...14S}
M.~{Sun}, J.~R. {Trump}, W.~N. {Brandt}, B.~{Luo}, D.~M. {Alexander},
  K.~{Jahnke}, D.~J. {Rosario}, S.~X. {Wang}, and Y.~Q. {Xue}, ApJ
  \textbf{802}, 14 (2015), arXiv: \eprint{1502.01025}.

\bibitem{Ding2020ApJ...888...37D}
X.~{Ding}, J.~{Silverman}, T.~{Treu}, A.~{Schulze}, M.~{Schramm}, S.~{Birrer},
  D.~{Park}, K.~{Jahnke}, V.~N. {Bennert}, J.~S. {Kartaltepe}, A.~M.
  {Koekemoer}, M.~A. {Malkan}, and D.~{Sanders}, ApJ \textbf{888}, 37 (2020),
  arXiv: \eprint{1910.11875}.

\bibitem{2007ApJ...670..249L}
T.~R. {Lauer}, S.~{Tremaine}, D.~{Richstone}, and S.~M. {Faber}, ApJ
  \textbf{670}, 249 (2007), arXiv: \eprint{0705.4103}.

\bibitem{2023arXiv230308150Z}
H.~{Zhang}, P.~{Behroozi}, M.~{Volonteri}, J.~{Silk}, X.~{Fan}, P.~F.
  {Hopkins}, J.~{Yang}, and J.~{Aird}, arXiv e-prints arXiv:2303.08150 (2023),
  arXiv: \eprint{2303.08150}.

\bibitem{2017MNRAS.472...90D}
X.~{Ding}, T.~{Treu}, S.~H. {Suyu}, K.~C. {Wong}, T.~{Morishita}, D.~{Park},
  D.~{Sluse}, M.~W. {Auger}, A.~{Agnello}, V.~N. {Bennert}, and T.~E.
  {Collett}, MNRAS \textbf{472}, 90 (2017), arXiv: \eprint{1703.02041}.

\bibitem{2023ApJ...947L..12R}
C.~L. {Rodriguez}, ApJl \textbf{947}, L12 (2023), arXiv: \eprint{2302.12386}.

\bibitem{2023arXiv230501307A}
R.~{Andrae} and K.~{El-Badry}, arXiv e-prints arXiv:2305.01307 (2023), arXiv:
  \eprint{2305.01307}.

\bibitem{2023arXiv230309391Y}
G.-W. {Yuan}, L.~{Lei}, Y.-Z. {Wang}, B.~{Wang}, Y.-Y. {Wang}, C.~{Chen}, Z.-Q.
  {Shen}, Y.-F. {Cai}, and Y.-Z. {Fan}, arXiv e-prints arXiv:2303.09391 (2023),
  arXiv: \eprint{2303.09391}.

\bibitem{2023arXiv230408153S}
T.~{Shinohara}, W.~{He}, Y.~{Matsuoka}, T.~{Nagao}, T.~{Suyama}, and
  T.~{Takahashi}, arXiv e-prints arXiv:2304.08153 (2023), arXiv:
  \eprint{2304.08153}.

\bibitem{2021IAUS..356..261T}
B.~{Trakhtenbrot}, in \emph{Nuclear Activity in Galaxies Across Cosmic Time},
  (edited by M.~{Povi{\'c}}, P.~{Marziani}, J.~{Masegosa}, H.~{Netzer}, S.~H.
  {Negu}, and S.~B. {Tessema}), volume 356, 261--275 (2021), arXiv:
  \eprint{2002.00972}.

\bibitem{2021ApJ...907....6Z}
P.~{Zhu}, L.~C. {Ho}, and H.~{Gao}, ApJ \textbf{907}, 6 (2021), arXiv:
  \eprint{2011.07216}.

\bibitem{2020NatAs...4..282S}
F.~{Shankar}, V.~{Allevato}, M.~{Bernardi}, C.~{Marsden}, A.~{Lapi},
  N.~{Menci}, P.~J. {Grylls}, M.~{Krumpe}, L.~{Zanisi}, F.~{Ricci}, F.~{La
  Franca}, R.~D. {Baldi}, J.~{Moreno}, and R.~K. {Sheth}, Nature Astronomy
  \textbf{4}, 282 (2020), arXiv: \eprint{1910.10175}.

\bibitem{2016ApJ...831..134V}
R.~C.~E. {van den Bosch}, ApJ \textbf{831}, 134 (2016), arXiv:
  \eprint{1606.01246}.

\bibitem{2023MNRAS.521.1023G}
A.~W. {Graham}, MNRAS \textbf{521}, 1023 (2023), arXiv: \eprint{2304.12524}.

\bibitem{2010A&A...522L...3S}
J.~E. {Sarria}, R.~{Maiolino}, F.~{La Franca}, F.~{Pozzi}, F.~{Fiore},
  A.~{Marconi}, C.~{Vignali}, and A.~{Comastri}, Astronomy and Astrophysics
  \textbf{522}, L3 (2010), arXiv: \eprint{1010.0768}.

\bibitem{2023A&A...672A..98M}
G.~{Mountrichas}, Astronomy and Astrophysics \textbf{672}, A98 (2023), arXiv:
  \eprint{2302.10937}.

\bibitem{Kent+etal+1994}
S.~M. {Kent}, Ap\&SS \textbf{217}, 27 (1994).

\bibitem{Fukugita+etal+1996}
M.~{Fukugita}, T.~{Ichikawa}, J.~E. {Gunn}, M.~{Doi}, K.~{Shimasaku}, and D.~P.
  {Schneider}, Astron. J. \textbf{111}, 1748 (1996).

\bibitem{Jedicke+etal+2007}
R.~{Jedicke} and {Pan-STARRS}, in \emph{AAS/Division for Planetary Sciences
  Meeting Abstracts \#39}, volume~39 of \emph{AAS/Division for Planetary
  Sciences Meeting Abstracts}, 8.02 (2007).

\bibitem{Chambers+etal+2016}
K.~C. {Chambers} and {Pan-STARRS Team}, in \emph{American Astronomical Society
  Meeting Abstracts \#227}, volume 227 of \emph{American Astronomical Society
  Meeting Abstracts}, 324.07 (2016).

\bibitem{Schmidt+etal+2005}
B.~P. {Schmidt}, S.~C. {Keller}, P.~J. {Francis}, and M.~S. {Bessell}, in
  \emph{American Astronomical Society Meeting Abstracts \#206}, volume 206 of
  \emph{American Astronomical Society Meeting Abstracts}, 15.09 (2005).

\bibitem{Rakich+etal+2006}
A.~{Rakich}, M.~{Blundell}, G.~{Pentland}, R.~{Brunswick}, T.~{Ferguson}, and
  J.~{Waltho}, in \emph{Society of Photo-Optical Instrumentation Engineers
  (SPIE) Conference Series}, (edited by L.~M. {Stepp}), volume 6267 of
  \emph{Society of Photo-Optical Instrumentation Engineers (SPIE) Conference
  Series}, 62670E (2006).

\bibitem{2005astro.ph.10346T}
{The Dark Energy Survey Collaboration}, arXiv e-prints astro-ph/0510346 (2005),
  arXiv: \eprint{astro-ph/0510346}.

\bibitem{2016MNRAS.460.1270D}
{Dark Energy Survey Collaboration}, T.~{Abbott}, F.~B. {Abdalla},
  J.~{Aleksi{\'c}}, S.~{Allam}, A.~{Amara}, D.~{Bacon}, E.~{Balbinot},
  M.~{Banerji}, K.~{Bechtol}, A.~{Benoit-L{\'e}vy}, G.~M. {Bernstein},
  E.~{Bertin}, J.~{Blazek}, C.~{Bonnett}, S.~{Bridle}, D.~{Brooks}, R.~J.
  {Brunner}, E.~{Buckley-Geer}, D.~L. {Burke}, G.~B. {Caminha}, D.~{Capozzi},
  J.~{Carlsen}, A.~{Carnero-Rosell}, M.~{Carollo}, M.~{Carrasco-Kind},
  J.~{Carretero}, F.~J. {Castander}, L.~{Clerkin}, T.~{Collett},
  C.~{Conselice}, M.~{Crocce}, C.~E. {Cunha}, C.~B. {D'Andrea}, L.~N. {da
  Costa}, T.~M. {Davis}, S.~{Desai}, H.~T. {Diehl}, J.~P. {Dietrich},
  S.~{Dodelson}, P.~{Doel}, A.~{Drlica-Wagner}, J.~{Estrada}, J.~{Etherington},
  A.~E. {Evrard}, J.~{Fabbri}, D.~A. {Finley}, B.~{Flaugher}, R.~J. {Foley},
  P.~{Fosalba}, J.~{Frieman}, J.~{Garc{\'\i}a-Bellido}, E.~{Gaztanaga}, D.~W.
  {Gerdes}, T.~{Giannantonio}, D.~A. {Goldstein}, D.~{Gruen}, R.~A. {Gruendl},
  P.~{Guarnieri}, G.~{Gutierrez}, W.~{Hartley}, K.~{Honscheid}, B.~{Jain},
  D.~J. {James}, T.~{Jeltema}, S.~{Jouvel}, R.~{Kessler}, A.~{King}, D.~{Kirk},
  R.~{Kron}, K.~{Kuehn}, N.~{Kuropatkin}, O.~{Lahav}, T.~S. {Li}, M.~{Lima},
  H.~{Lin}, M.~A.~G. {Maia}, M.~{Makler}, M.~{Manera}, C.~{Maraston}, J.~L.
  {Marshall}, P.~{Martini}, R.~G. {McMahon}, P.~{Melchior}, A.~{Merson}, C.~J.
  {Miller}, R.~{Miquel}, J.~J. {Mohr}, X.~{Morice-Atkinson}, K.~{Naidoo},
  E.~{Neilsen}, R.~C. {Nichol}, B.~{Nord}, R.~{Ogando}, F.~{Ostrovski},
  A.~{Palmese}, A.~{Papadopoulos}, H.~V. {Peiris}, J.~{Peoples}, W.~J.
  {Percival}, A.~A. {Plazas}, S.~L. {Reed}, A.~{Refregier}, A.~K. {Romer},
  A.~{Roodman}, A.~{Ross}, E.~{Rozo}, E.~S. {Rykoff}, I.~{Sadeh}, M.~{Sako},
  C.~{S{\'a}nchez}, E.~{Sanchez}, B.~{Santiago}, V.~{Scarpine}, M.~{Schubnell},
  I.~{Sevilla-Noarbe}, E.~{Sheldon}, M.~{Smith}, R.~C. {Smith},
  M.~{Soares-Santos}, F.~{Sobreira}, M.~{Soumagnac}, E.~{Suchyta},
  M.~{Sullivan}, M.~{Swanson}, G.~{Tarle}, J.~{Thaler}, D.~{Thomas}, R.~C.
  {Thomas}, D.~{Tucker}, J.~D. {Vieira}, V.~{Vikram}, A.~R. {Walker}, R.~H.
  {Wechsler}, J.~{Weller}, W.~{Wester}, L.~{Whiteway}, H.~{Wilcox}, B.~{Yanny},
  Y.~{Zhang}, and J.~{Zuntz}, MNRAS \textbf{460}, 1270 (2016), arXiv:
  \eprint{1601.00329}.

\bibitem{Hlozek+etal+2019}
R.~{Hlozek}, J.~{Albert}, M.~{Balogh}, P.~{Barmby}, J.~{Blakeslee}, J.~{Bovy},
  P.~{Cote}, S.~{Cote}, J.~{Di Francesco}, M.~{Drout}, G.~{Eadie}, S.~{Fabbro},
  W.~{Fraser}, B.~{Gaensler}, S.~{Gallagher}, M.~{Graham}, D.~{Haggard},
  P.~{Hall}, C.~{Heinke}, M.~{Hudson}, J.~{Hutchings}, J.~{Kavelaars},
  S.~{Lawler}, D.~{Leahy}, A.~{McConnachie}, W.~{Percival}, C.~{Pritchet},
  M.~{Rahman}, J.~{Ruan}, M.~{Sawicki}, G.~{Sivakoff}, J.~{Taylor},
  K.~{Thanjavur}, and P.~{Wiegert}, in \emph{Canadian Long Range Plan for
  Astronomy and Astrophysics White Papers}, volume 2020, 51 (2019).

\bibitem{Lou+etal+2016}
Z.~{Lou}, M.~{Liang}, D.~{Yao}, X.~{Zheng}, J.~{Cheng}, H.~{Wang}, W.~{Liu},
  Y.~{Qian}, H.~{Zhao}, and J.~{Yang}, in \emph{Society of Photo-Optical
  Instrumentation Engineers (SPIE) Conference Series}, volume 10154 of
  \emph{Society of Photo-Optical Instrumentation Engineers (SPIE) Conference
  Series}, 101542A (2016).

\bibitem{Lei+etal+2023}
L.~{Lei}, Q.-F. {Zhu}, X.~{Kong}, T.-G. {Wang}, X.-Z. {Zheng}, D.-D. {Shi},
  L.-L. {Fan}, and W.~{Liu}, Research in Astronomy and Astrophysics
  \textbf{23}, 035013 (2023), arXiv: \eprint{2301.03068}.

\bibitem{WFSTscience}
{WFST Collaboration}, T.~{Wang}, G.~{Liu}, Z.~{Cai}, J.~{Geng}, M.~{Fang},
  H.~{He}, J.-a. {Jiang}, N.~{Jiang}, X.~{Kong}, B.~{Li}, Y.~{Li}, W.~{Luo},
  Z.~{Pan}, X.~{Wu}, J.~{Yang}, J.~{Yu}, X.~{Zheng}, Q.~{Zhu}, Y.-F. {Cai},
  Y.~{Chen}, Z.~{Chen}, Z.~{Dai}, L.~{Fan}, Y.~{Fan}, W.~{Fang}, Z.~{He},
  L.~{Hu}, M.~{Hu}, Z.~{Jin}, Z.~{Jiang}, G.~{Li}, F.~{Li}, X.~{Li},
  R.~{Liang}, Z.~{Lin}, Q.~{Liu}, W.~{Liu}, Z.~{Liu}, W.~{Liu}, Y.~{Liu},
  Z.~{Lou}, H.~{Qu}, Z.~{Sheng}, J.~{Shi}, Y.~{Shu}, Z.~{Su}, T.~{Sun},
  H.~{Wang}, H.~{Wang}, J.~{Wang}, J.~{Wang}, D.~{Wei}, J.~{Wei}, Y.~{Xue},
  J.~{Yan}, C.~{Yang}, Y.~{Yuan}, Y.~{Yuan}, H.~{Zhang}, M.~{Zhang}, H.~{Zhao},
  and W.~{Zhao}, SCIENCE CHINA Physics, Mechanics \& Astronomy \textbf{66},
  109512 (2023), arXiv: \eprint{2306.07590},
  \urlprefix\url{https://www.sciengine.com/SCPMA/doi/10.1007/s11433-023-2197-5}.

\bibitem{Liu+etal+2019}
X.~{Liu}, in \emph{Galactic Archaeology in the Gaia Era}, 14 (2019).

\bibitem{Lei+etal+2020}
L.~{Lei}, J.~D. {Li}, J.~T. {Wu}, S.~Y. {Jiang}, and B.~Q. {Chen}, Astronomical
  Research \& Technology \textbf{18}, 115121 (2021), arXiv:
  \eprint{202007.00022}, \urlprefix\url{http://chinaxiv.org/abs/202007.00022}.

\bibitem{Yuan+etal+2020}
X.~{Yuan}, Z.~{Li}, X.~{Liu}, D.~{Niu}, Q.~{Lu}, F.~{Jiang}, Y.~{Wang},
  X.~{Li}, Y.~{Liang}, H.~{Wang}, C.~{Zhang}, J.~{Wang}, B.~{Li}, J.~{Tian},
  H.~{Lu}, B.~{Chen}, Y.~{Huang}, X.~{Liu}, Z.~{Yao}, X.~{Cui}, and G.~{Li}, in
  \emph{Society of Photo-Optical Instrumentation Engineers (SPIE) Conference
  Series}, volume 11445 of \emph{Society of Photo-Optical Instrumentation
  Engineers (SPIE) Conference Series}, 114457M (2020).

\bibitem{Lei+etal+2022}
L.~{Lei}, B.-Q. {Chen}, J.-D. {Li}, J.-T. {Wu}, S.-Y. {Jiang}, and X.-W. {Liu},
  Research in Astronomy and Astrophysics \textbf{22}, 025004 (2022), arXiv:
  \eprint{2111.08316}.

\bibitem{Zhao+etal+2016}
H.~{Zhao}, Y.~{Li}, and C.~{Zhang}, IEEE Geoscience and Remote Sensing Letters
  \textbf{13}, 1139 (2016).

\bibitem{Yuan+etal+2021}
H.-B. {Yuan}, D.-S. {Deng}, and Y.~{Sun}, Research in Astronomy and
  Astrophysics \textbf{21}, 074 (2021), arXiv: \eprint{2010.14005}.

\bibitem{McEnery+etal+2021}
J.~{McEnery}, in \emph{American Astronomical Society Meeting Abstracts},
  volume~53 of \emph{American Astronomical Society Meeting Abstracts}, 327.01
  (2021).

\bibitem{2006SPIE.6266E..0PP}
P.~{Predehl}, G.~{Hasinger}, H.~{B{\"o}hringer}, U.~{Briel}, H.~{Brunner},
  E.~{Churazov}, M.~{Freyberg}, P.~{Friedrich}, E.~{Kendziorra}, D.~{Lutz},
  N.~{Meidinger}, M.~{Pavlinsky}, E.~{Pfeffermann}, A.~{Santangelo},
  J.~{Schmitt}, P.~{Schuecker}, A.~{Schwope}, M.~{Steinmetz}, L.~{Str{\"u}der},
  R.~{Sunyaev}, and J.~{Wilms}, in \emph{Space Telescopes and Instrumentation
  II: Ultraviolet to Gamma Ray}, (edited by M.~J.~L. {Turner} and
  G.~{Hasinger}), volume 6266 of \emph{Society of Photo-Optical Instrumentation
  Engineers (SPIE) Conference Series}, 62660P (2006).

\bibitem{2006SPIE.6266E..0OP}
M.~{Pavlinsky}, G.~{Hasinger}, A.~{Parmar}, G.~{Fraser}, E.~{Churazov},
  M.~{Gilfanov}, R.~{Sunyaev}, A.~{Vikhlinin}, P.~{Predehl}, L.~{Piro},
  V.~{Arefiev}, A.~{Tkachenko}, V.~{Pinchuk}, and D.~{Gorobets}, in \emph{Space
  Telescopes and Instrumentation II: Ultraviolet to Gamma Ray}, (edited by
  M.~J.~L. {Turner} and G.~{Hasinger}), volume 6266 of \emph{Society of
  Photo-Optical Instrumentation Engineers (SPIE) Conference Series}, 62660O
  (2006).

\bibitem{2022hxga.book...86Y}
W.~{Yuan}, C.~{Zhang}, Y.~{Chen}, and Z.~{Ling}, in \emph{Handbook of X-ray and
  Gamma-ray Astrophysics. Edited by Cosimo Bambi and Andrea Santangelo}, 86
  (2022).

\bibitem{2015arXiv150607735Y}
W.~{Yuan}, C.~{Zhang}, H.~{Feng}, S.~N. {Zhang}, Z.~X. {Ling}, D.~{Zhao},
  J.~{Deng}, Y.~{Qiu}, J.~P. {Osborne}, P.~{O'Brien}, R.~{Willingale},
  J.~{Lapington}, G.~W. {Fraser}, and {the Einstein Probe team}, arXiv e-prints
  arXiv:1506.07735 (2015), arXiv: \eprint{1506.07735}.

\bibitem{2013A&A...558A..33A}
{Astropy Collaboration}, T.~P. {Robitaille}, E.~J. {Tollerud}, P.~{Greenfield},
  M.~{Droettboom}, E.~{Bray}, T.~{Aldcroft}, M.~{Davis}, A.~{Ginsburg}, A.~M.
  {Price-Whelan}, W.~E. {Kerzendorf}, A.~{Conley}, N.~{Crighton}, K.~{Barbary},
  D.~{Muna}, H.~{Ferguson}, F.~{Grollier}, M.~M. {Parikh}, P.~H. {Nair}, H.~M.
  {Unther}, C.~{Deil}, J.~{Woillez}, S.~{Conseil}, R.~{Kramer}, J.~E.~H.
  {Turner}, L.~{Singer}, R.~{Fox}, B.~A. {Weaver}, V.~{Zabalza}, Z.~I.
  {Edwards}, K.~{Azalee Bostroem}, D.~J. {Burke}, A.~R. {Casey}, S.~M.
  {Crawford}, N.~{Dencheva}, J.~{Ely}, T.~{Jenness}, K.~{Labrie}, P.~L. {Lim},
  F.~{Pierfederici}, A.~{Pontzen}, A.~{Ptak}, B.~{Refsdal}, M.~{Servillat}, and
  O.~{Streicher}, Astronomy and Astrophysics \textbf{558}, A33 (2013), arXiv:
  \eprint{1307.6212}.

\bibitem{2018AJ....156..123A}
{Astropy Collaboration}, A.~M. {Price-Whelan}, B.~M. {Sip{\H{o}}cz}, H.~M.
  {G{\"u}nther}, P.~L. {Lim}, S.~M. {Crawford}, S.~{Conseil}, D.~L. {Shupe},
  M.~W. {Craig}, N.~{Dencheva}, A.~{Ginsburg}, J.~T. {VanderPlas}, L.~D.
  {Bradley}, D.~{P{\'e}rez-Su{\'a}rez}, M.~{de Val-Borro}, T.~L. {Aldcroft},
  K.~L. {Cruz}, T.~P. {Robitaille}, E.~J. {Tollerud}, C.~{Ardelean},
  T.~{Babej}, Y.~P. {Bach}, M.~{Bachetti}, A.~V. {Bakanov}, S.~P. {Bamford},
  G.~{Barentsen}, P.~{Barmby}, A.~{Baumbach}, K.~L. {Berry}, F.~{Biscani},
  M.~{Boquien}, K.~A. {Bostroem}, L.~G. {Bouma}, G.~B. {Brammer}, E.~M. {Bray},
  H.~{Breytenbach}, H.~{Buddelmeijer}, D.~J. {Burke}, G.~{Calderone}, J.~L.
  {Cano Rodr{\'\i}guez}, M.~{Cara}, J.~V.~M. {Cardoso}, S.~{Cheedella},
  Y.~{Copin}, L.~{Corrales}, D.~{Crichton}, D.~{D'Avella}, C.~{Deil},
  {\'E}.~{Depagne}, J.~P. {Dietrich}, A.~{Donath}, M.~{Droettboom}, N.~{Earl},
  T.~{Erben}, S.~{Fabbro}, L.~A. {Ferreira}, T.~{Finethy}, R.~T. {Fox}, L.~H.
  {Garrison}, S.~L.~J. {Gibbons}, D.~A. {Goldstein}, R.~{Gommers}, J.~P.
  {Greco}, P.~{Greenfield}, A.~M. {Groener}, F.~{Grollier}, A.~{Hagen},
  P.~{Hirst}, D.~{Homeier}, A.~J. {Horton}, G.~{Hosseinzadeh}, L.~{Hu}, J.~S.
  {Hunkeler}, {\v{Z}}.~{Ivezi{\'c}}, A.~{Jain}, T.~{Jenness}, G.~{Kanarek},
  S.~{Kendrew}, N.~S. {Kern}, W.~E. {Kerzendorf}, A.~{Khvalko}, J.~{King},
  D.~{Kirkby}, A.~M. {Kulkarni}, A.~{Kumar}, A.~{Lee}, D.~{Lenz}, S.~P.
  {Littlefair}, Z.~{Ma}, D.~M. {Macleod}, M.~{Mastropietro}, C.~{McCully},
  S.~{Montagnac}, B.~M. {Morris}, M.~{Mueller}, S.~J. {Mumford}, D.~{Muna},
  N.~A. {Murphy}, S.~{Nelson}, G.~H. {Nguyen}, J.~P. {Ninan}, M.~{N{\"o}the},
  S.~{Ogaz}, S.~{Oh}, J.~K. {Parejko}, N.~{Parley}, S.~{Pascual}, R.~{Patil},
  A.~A. {Patil}, A.~L. {Plunkett}, J.~X. {Prochaska}, T.~{Rastogi}, V.~{Reddy
  Janga}, J.~{Sabater}, P.~{Sakurikar}, M.~{Seifert}, L.~E. {Sherbert},
  H.~{Sherwood-Taylor}, A.~Y. {Shih}, J.~{Sick}, M.~T. {Silbiger},
  S.~{Singanamalla}, L.~P. {Singer}, P.~H. {Sladen}, K.~A. {Sooley},
  S.~{Sornarajah}, O.~{Streicher}, P.~{Teuben}, S.~W. {Thomas}, G.~R.
  {Tremblay}, J.~E.~H. {Turner}, V.~{Terr{\'o}n}, M.~H. {van Kerkwijk}, A.~{de
  la Vega}, L.~L. {Watkins}, B.~A. {Weaver}, J.~B. {Whitmore}, J.~{Woillez},
  V.~{Zabalza}, and {Astropy Contributors}, Astron. J. \textbf{156}, 123
  (2018), arXiv: \eprint{1801.02634}.

\bibitem{2013PASP..125..306F}
D.~{Foreman-Mackey}, D.~W. {Hogg}, D.~{Lang}, and J.~{Goodman}, PASP
  \textbf{125}, 306 (2013), arXiv: \eprint{1202.3665}.

\bibitem{Hunter:2007}
J.~D. Hunter, Computing in Science \& Engineering \textbf{9}, 90 (2007).

\bibitem{harris2020array}
C.~R. Harris, K.~J. Millman, S.~J. van~der Walt, R.~Gommers, P.~Virtanen,
  D.~Cournapeau, E.~Wieser, J.~Taylor, S.~Berg, N.~J. Smith, R.~Kern, M.~Picus,
  S.~Hoyer, M.~H. van Kerkwijk, M.~Brett, A.~Haldane, J.~F. del R{\'{i}}o,
  M.~Wiebe, P.~Peterson, P.~G{\'{e}}rard-Marchant, K.~Sheppard, T.~Reddy,
  W.~Weckesser, H.~Abbasi, C.~Gohlke, and T.~E. Oliphant, Nature \textbf{585},
  357 (2020), \urlprefix\url{https://doi.org/10.1038/s41586-020-2649-2}.

\end{thebibliography}








\end{multicols}
\end{document}